\documentclass[final]{agujournal2019}
\usepackage{url} 
\usepackage{soul}
\usepackage{gensymb}
\usepackage{amsmath,amsfonts,amsthm}
\usepackage{graphics, graphicx, caption}
\usepackage{amsmath}

\journalname{Space Weather}

\begin{document}

%
%

\twocolumn[
\begin{@twocolumnfalse}
\title{Resolving moving heliospheric structures using interplanetary scintillation observations with the Murchison Widefield Array}

%




\authors{A. Waszewski\affil{1,3}, J. S. Morgan\affil{1}, R. Chhetri\affil{2}, R. Ekers\affil{1,3}, M. C. M. Cheung\affil{3}, N. D. R. Bhat\affil{1}, M.~Johnston-Hollitt\affil{4}}


\affiliation{1}{International Centre for Radio Astronomy Research, Curtin University, Bentley, WA 6102, Australia}
\affiliation{2}{CSIRO Space and Astronomy, P.O. Box 1130, Bentley, WA 6102, Australia}
\affiliation{3}{CSIRO Space and Astronomy, P.O. Box 76, Epping, NSW 1710, Australia}
\affiliation{4}{Curtin Institute for Data Science, Curtin University, Bentley, WA 6102, Australia}




\correspondingauthor{Angelica Waszewski}{angelica.waszewski@icrar.org}




\begin{keypoints}
\item We conducted a blind search for heliospheric activity in 93 MWA IPS observations spanning 49 days, which contain a high density of sources.
\item We have made the first unambiguous detection of a heliospheric structure using IPS measurements but find no coronagraph counterpart.
\item From two independent IPS observations we have inferred a plane-of-sky angular velocity within an error of 22\%.
\end{keypoints}

%
%

%
%


\begin{abstract}
We have conducted a blind search in 49 consecutive days of interplanetary scintillation observations made by the Murchison Widefield Array from mid-2019, with overlapping daily observations approximately East and South-East of the Sun at an elongation of $\sim$30 degrees and a field of view of 30 degrees. These observations detect an unprecedented density of sources. In spite of these observations being taken at sunspot minimum, this search has revealed several interesting transitory features characterised by elevated scintillation levels.
One solar wind enhancement is captured in two observations several hours apart, allowing its radial movement away from the Sun to be measured. We present here a methodology for measuring the plane-of-sky velocity for the moving heliospheric structure. The plane-of-sky velocity was inferred as $0.66\pm0.147\,^{\text{o}}\text{hr}^{-1}$, or $480\pm106\,\text{km}\,\text{s}^{-1}$\, assuming a distance of 1AU. 
After cross-referencing our observed structure with multiple catalogues of heliospheric events, we propose that the likely source of our observed structure is a stream-interaction region originating from a low-latitude coronal hole. 
This work demonstrates the power of widefield interplanetary scintillation observations to capture detailed features in the heliosphere which are otherwise unresolvable and go undetected. 
\end{abstract}

\section*{Plain Language Summary}
We used the Murchison Widefield Array to observe the heliosphere for 49 days and we found several interesting transitory features, including a solar wind enhancement that was captured in two observations several hours apart. This allowed us to measure the radial movement of the solar wind enhancement away from the Sun, at a velocity of $480\pm106\,\text{km}\,\text{s}^{-1}$\, assuming a distance of 1AU. We have also determined that the likely source of the solar wind enhancement is a stream-interaction region. This work demonstrates the power of widefield interplanetary scintillation observations to capture detailed features in the heliosphere which are otherwise unresolvable and go undetected.
\newline
\end{@twocolumnfalse}
]

%
%






\section{Introduction}
\label{sect:intro}
Interplanetary scintillation (IPS) was discovered in 1964 \cite{clarke, HEWISH1964}. It was observed as amplitude scintillation at radio wavelengths as radiation from compact objects (0.3 arcsec for 162\,MHz) traverse through the irregularities of the solar wind \cite{Coles1978}. IPS, as a general technique, has been used to study the solar wind, solar wind transients, and for inner-heliospheric observations for over 55 years. This technique can give exclusive perspectives of the behaviour of the heliosphere. In particular, IPS observations allow for the solar wind to be inferred over all solar latitudes and a wide range of heliocentric distances \cite<e.g.>{Bisi2009}. IPS is often characterised by the g-level \cite<e.g., >{Gapper1982}, also known as the scintillation enhancement factor, which is the simplest measure of space weather. As g-level is dependent on scintillation, it can be used as a proxy for the density of the solar wind along the line-of-sight \cite{Tappin1986}. As is common with space weather analysis using IPS, the g-level measure is widely used in this paper (e.g., Section\,\ref{sect:g-level} and following sections).

Recent studies that use IPS techniques for solar weather purposes \cite<e.g.>{Iwai2019, Iwai2021, Tokumaru2000, TOKUMARU2013, Kojima2013, Chang2016, Jackson2007, Jackson2008, Tokumaru2019} all use instruments such as the Solar Wind Imaging Facility \cite<SWIFT,>{Tokumaru2011}, the Ooty Radio Telescope \cite<ORT,>{Manoharan2010}, the Low-Frequency Array \cite<LOFAR,>{lofar}, and the Mexican Array Radio Telescope \cite<MEXART,>{Esparza2004}. The Murchison Widefield Array \cite<MWA,>[]{Tingay, Wayth2018}, brings unique capabilities to the field as it can be used to make IPS measurements of hundreds of sources across the southern sky simultaneously, owing to its large field of view ($\sim900\,\text{deg}^2$ at 160\,MHz) and higher instantaneous sensitivity.

The MWA is a low-frequency radio telescope, operating in the frequency range of 70-300\,MHz and is located in the Murchison shire of Western Australia. It consists of 4\,096 antennas arranged in 256 tiles (128 of which are used at any one time) of 4$\times$4 dipoles, which are distributed over an area extending over several kilometres in radius. With longer baselines extending out to $\sim$5-6\,km, the MWA provides an angular resolution $\sim$1 arc-minute (at 160MHz). The MWA was designed and built to be a flexible general purpose instrument, supporting many different science goals \cite{Beardsley2019}. The MWA is able to provide a time resolution of 0.5\,s in imaging mode, short enough to measure IPS, allowing \citeA{morgan2018} and \citeA{Chhetri2018} to use IPS in order to identify and characterise compact radio sources. At MWA frequencies, this can most easily be done at solar elongations of around 30$^\circ$. More recently, \citeA{Morgan2023} demonstrated that a Coronal Mass Ejection (CME), detected at launch in white-light coronagraph images, could be detected in interplanetary space using MWA IPS observations, with the unprecedented number of lines of sight allowing the CME to be mapped in exquisite detail, using just two 5-minute observations.

More recent data, taken by the upgraded Phase II MWA \cite{Wayth2018} has been synthesised by \citeA{ipssurvey} into the first data release of the Phase II IPS survey, which catalogs the IPS properties of over 40\,000 sources from the GLEAM survey \cite{Wayth2015, hurley-walker2017}, with IPS strongly detected in over 7000 of them. As well as providing baseline scintillation levels for all of these sources, facilitating space weather studies, a further byproduct of this survey is a set of 250 10-minute MWA IPS observations, with the scintillation index of several hundred IPS sources measured in each of them. These data are the starting point for the study presented here.

A subset of 93 observations taken across 49 contiguous days of observations was chosen from the survey to form the basis of this study, since they provide continuous observations of the eastern limb of the Sun. While conducting a blind search of this data, we identified several heliospheric structures. One in particular was detected in two observations spaced 96 minutes apart, allowing us to infer the plane-of-sky angular velocity of the structure.
This structure is the focus of this paper, which is organised as follows: in Section\,\ref{sect:method} we describe how the MWA IPS observations were chosen, and how they were further reduced to probe the solar wind. Section\,\ref{sect:result} outlines the identification of the heliospheric structure, and discusses the novel method of inferring the plane-of-sky velocity with errors, calculated using jack-knife tests \cite{jackknife}, of a solar wind structure in IPS observations. We also discuss the possible solar origins of this structure. Section\,\ref{sect:discuss} discusses the implications that this work has on future heliospheric studies and modeling, and concludes with suggestions for further work. To describe our observations we use a range of heliocentric based coordinate systems. A description of each frame can be found in Section\,\ref{sect:def}.


\section{Methodology}
\label{sect:method}
\subsection{Observations}
\label{sect:observations}
\begin{figure*}
    \centering
    \includegraphics[width=\textwidth]{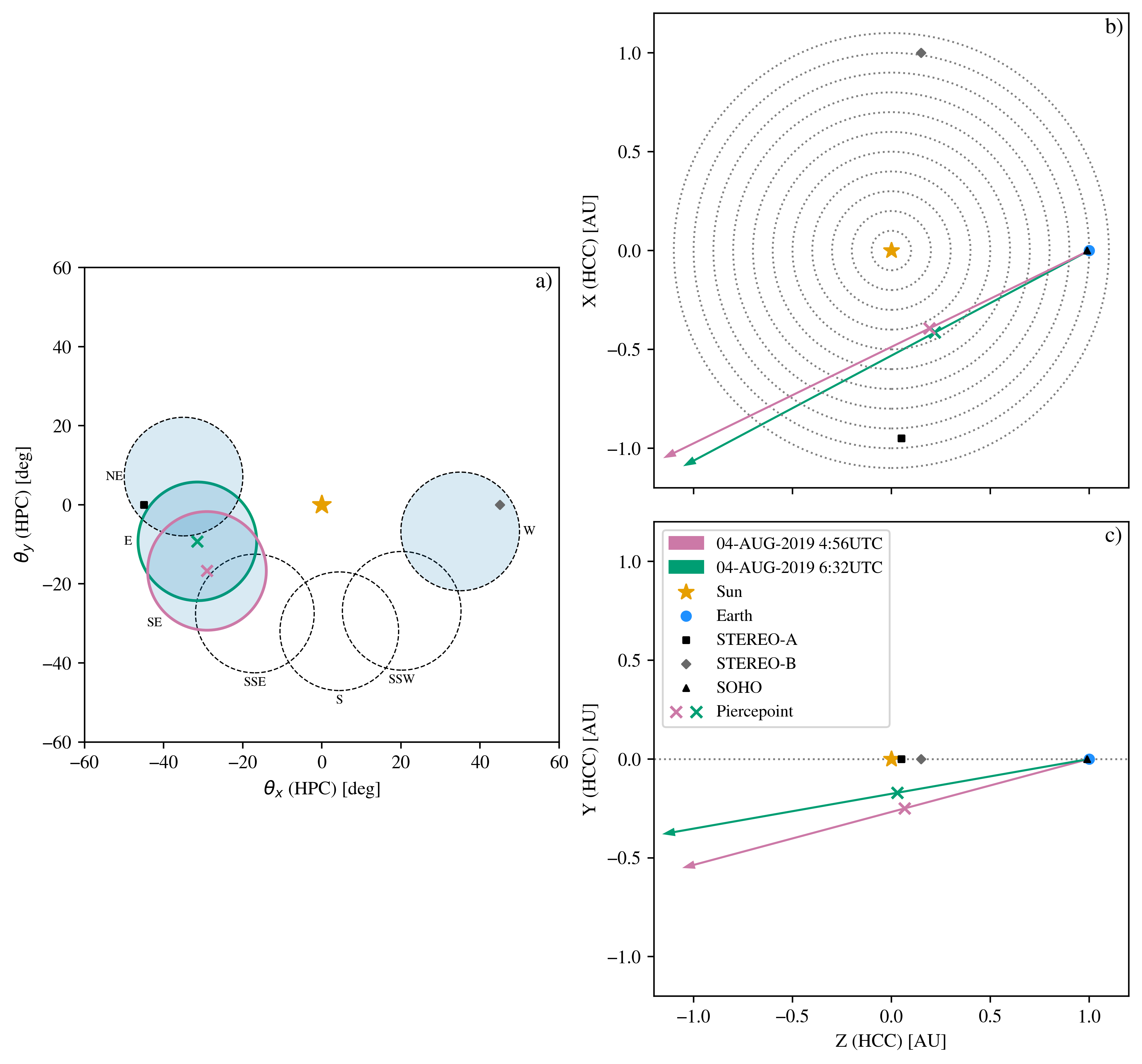}
    \caption{
    a) IPS survey target fields for 2019-AUG-04, including 4 of the principal target fields coloured in blue (W, $90^\text{o}$; SE, $240^\text{o}$; E, $270^\text{o}$; NE, $300^\text{o}$; all relative to Ecliptic North) and 3 of the additional target fields only outlined (SSW, $150^\text{o}$; S, $180^\text{o}$; SSE, $210^\text{o}$) in helioprojective coordinates (HPC). All pointings are taken at $30^\text{o}$ from the Sun (0.5\,AU or $\sim\,107$ solar radii). The two pointing directions chosen for the blind search are outlined in a solid line, with pink for the SE (earlier observation) and green for the E (later observation) pointings. The locations of STEREO-A (black square) and STEREO-B (grey diamond) are also shown.
    b) The line of sight for both target pointings in the ecliptic plane, with their associated piercepoints (point of closest approach with the Sun) in heliocentric coordinates (HCC). The locations of STEREO-A (black square), STEREO-B (grey diamond), and SOHO (black triangle) are also shown.
    c) The line of sight for both target pointings in the meridional plane, with their associated piercepoints in HCC. The locations of STEREO-A, STEREO-B, and SOHO (black triangle) are also shown.}
    \label{fig:target}
\end{figure*}

For the MWA Phase II IPS survey, 10-minute observations were taken almost daily between 2019-FEB-04 and 2019-AUG-18, totalling in 1\,511 observations. The survey sampled from 6 principal target fields at a $\epsilon$ of $30^\text{o}$ ($\sim107$ solar radii from the Sun) at $\phi$ of $60^\text{o}$, $90^\text{o}$, $120^\text{o}$, $240^\text{o}$, $270^\text{o}$, $300^\text{o}$, with an occasional additional 5 target fields at $30^\text{o}$, $150^\text{o}$, $180^\text{o}$, $210^\text{o}$, $330^\text{o}$ \cite{ipssurvey}; see left-most panel of Fig.\,\ref{fig:target}. However, only 263 of the total observations were selected for the first data release \cite{ipssurvey}, a large number of observations remain unprocessed. For the first blind search for space weather events we chose to focus on a set of 93 observation spanning over 49 contiguous days (spanning from 2019-JUN-25 to 2019-AUG-14, with two days with no observations taken, 2019-JUL-16 and 2019-AUG-01), with two pointing directions (E and SE, specified with a solid line outline in Fig.\,\ref{fig:target}) observed on all but five days (2019-JUN-25, 2019-JUN-29, 2019-JUL-02, 2019-JUL-27, and 2019-AUG-06) which only had the SE pointing direction processed.  

\subsection{Space Weather Analysis of MWA Data}
\label{sect:g-level}
In the radio wavelengths there are other sources of variability that can influence how a source scintillates. One of the main contributors is ionospheric effects. Ionospheric scintillation acts on a longer time-scale and also affects sources that are much larger in size. These differences allow ionospheric scintillation to be easily filtered out in the variability of compact radio sources \cite{morgan2018}. The full filtering, calibration \cite{offringa-2015-mwa-environment}, and imaging procedure, as well as methodology for extracting variability from the image plane, are outlined in detail in Section\,2 and 3 of \citeA{ipssurvey}. Here we describe the extra steps needed for a space weather analysis of the same data. Note that a space weather analysis of MWA IPS data containing a CME has already been carried out \cite{Morgan2023}, but the approach used here differs in that we used the published catalogue of \citeA{ipssurvey} to provide the reference scintillation levels required. 

\citeA{ipssurvey} model the scintillation index, $m_{\text{pt}}$, of an unresolved source due to the average background solar wind as
\begin{equation}
    m_{\text{pt}} = 0.06\lambda(ep)^{-b},
    \label{eq:point}
\end{equation}
\cite{Rickett1973, Manoharan1993}, where $\lambda$ is the wavelength (in metres) of the observation, $e$ is the elliptical term defined as
\begin{equation}
    e = \sqrt{2.25\sin^2(\phi) + \cos^2(\phi)}
    \label{eq:ellip}
\end{equation}
\cite{Morgan2019}, and $p$ is the point of closest approach of the life-of-sight to the Sun, also known as the piercepoint. At a solar elongation of $\epsilon$, the piercepoint is at a distance of sin($\epsilon$) (in AU) from the Sun. The geometry of the IPS line-of-sight for the chosen pointings in this work is shown in the right-hand panels of Fig\,\ref{fig:target}, showing where the piercepoint is located depending on the pointing around the Sun looking at both the ecliptic and meridional plane. The constant of proportionality for $m_{\text{pt}}$, the ellipticity term $e$, and the power-law index $b$, will all vary during the solar cycle, as well as between solar cycles. Where applicable we have used their long term average values for solar minimum which have been established as a constant of proportionality of 0.06 and a power law index, $b \approx 1.6$ \cite[ and references therein]{ipssurvey, Morgan2019}.
We use as our starting point the scintillation indices, error estimations, and normalised scintillation indices \cite<NSI,>{Chhetri2018} for each source in each observation from the processed observations used in \citeA{ipssurvey}. Sources in general vary in scintillation level due to the inherent source structure, therefore we use the NSI per source as it gives the scintillation of a source relative to a compact source. 

\subsubsection{Determining g-levels}
The g-level, as introduced in Section\,\ref{sect:intro}, also known as the scintillation enhancement factor, is a measure of how much a particular source's scintillation is departing from its norm in a particular observation. In basic terms it is the observed scintillation level relative to a baseline, expected scintillation and is defined as, 
\begin{linenomath*}
 \begin{equation}
    g = \frac{m_{\rm{obs}}}{m_{\rm{pt}} \times \rm{NSI}},
 \end{equation}
 \end{linenomath*}
where $m_{\rm{obs}}$ is the observed scintillation of a source in a particular observation. 
As mentioned previously in Section\,\ref{sect:intro}, the g-level is used as a proxy for the density of the solar wind along the line-of-sight \cite{Tappin1986} as it is dependent on scintillation. The scintillation level is approximately proportional to the square of the electron density integrated along the line of sight \cite[ and references therein]{Morgan2019}, but it should be noted that the exact relationship between the g-level and the electron density is not relevant to the analysis that is done in this work. We defer the determination of the quantitative physical characteristic to future work. However, for the purposes of detecting structures in the g-level, a proper baseline scintillation level must be used. It is necessary to define a baseline scintillation level which takes into account the distance from the Sun, as well as the fact that the polar solar wind is more diffuse. The ellipticity term defined in Eq.\,\ref{eq:ellip} will account for the latter, therefore we combine the expected scintillation of a source from Eq.\,\ref{eq:point} with the NSI to account for source structure, to give the expected scintillation level of the source in question. This differs to the analysis in \citeA{Morgan2023}, as the NSI for each source was calculated over the full set of observations taken for the survey, rather than just using the observation in question. 

\begin{figure*}[h]
    \centering
    \includegraphics[width=\textwidth]{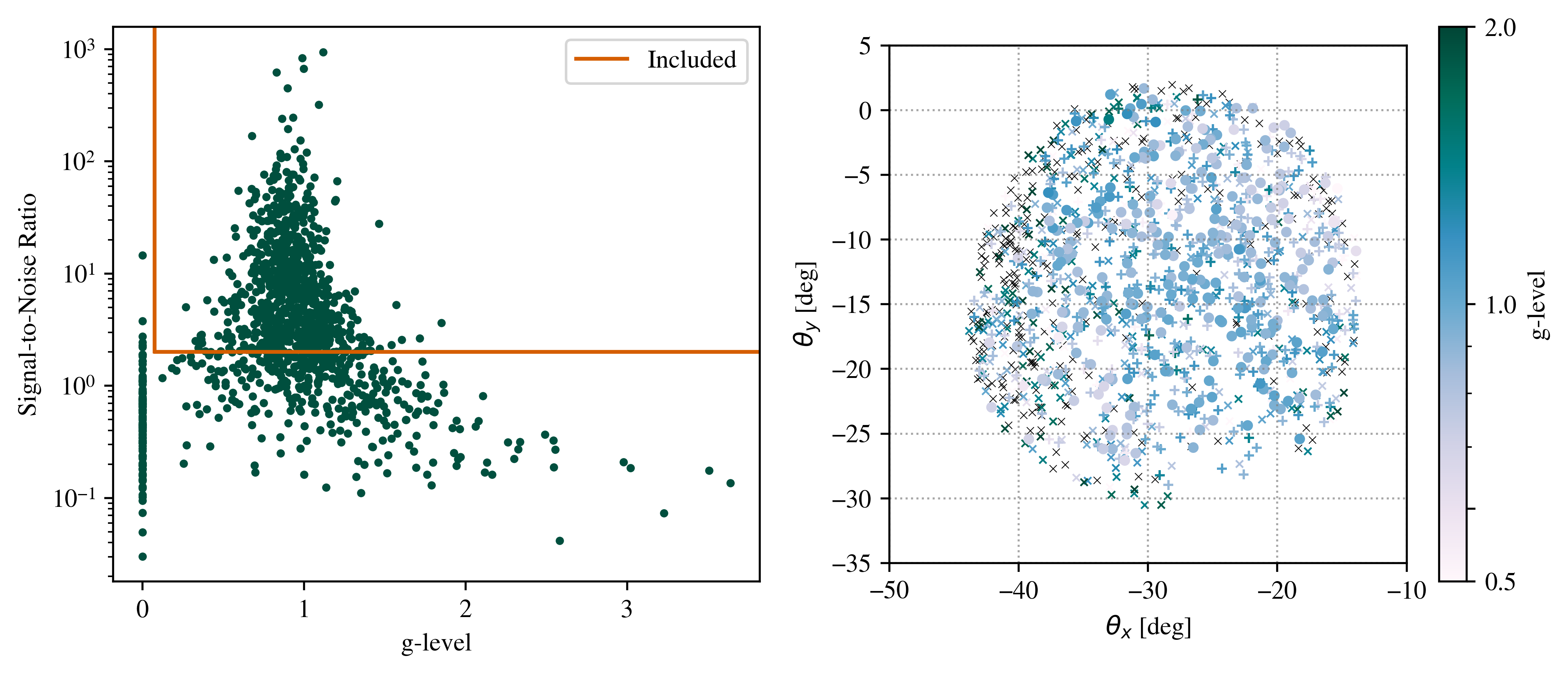}
    \caption{Left: Distribution of g-level compared to a source's signal to noise (S/N) within a MWA IPS observation. All sources on this distribution are included in the g-map on the left, but for further analysis, only those above the orange line are included. These sources meet the criteria of having a defined g-level with a S/N of above 2.
    Right: A heliocentric (Sun at the origin) g-map of a MWA IPS observation with 1\,345 sources, where 78\% of sources within the field have an associated g-level, either shown as a coloured circle (S/N of above 5) or a coloured plus (S/N of above 2, but below 5). Sources with no defined g-level are shown as black crosses, and those that have a defined g-level but are below a S/N of 2 are a coloured cross (colour associated to its g-level). These sources shown in crosses are excluded in further analysis.
    This example observation was taken on 2019-JUL-21 05:30 UTC.
    }
    \label{fig:random}
\end{figure*}

By calculating the g-level of every source within the field of view (FOV), we can add this information to a map of the sky called a g-map, which can be used to detect regions of enhanced scintillation. An example is shown on the right hand panel of Fig.\,\ref{fig:random}. In this particular observation, there are $1\,345$ sources, $1\,055$ have measured g-levels. The majority of sources in the g-map in Fig.\,\ref{fig:random} are clustered around a g-level of 1, as shown by the distribution of g-levels to the left of Fig.\,\ref{fig:random}, with only a handful showing major deviations in scintillation levels. At the high signal to noise limit there is a scatter of about 20\% around a g-level of 1, consistent with what is shown in \citeA{Tappin1986}.
Therefore we conclude that in this case there is less large-scale structure (fewer to no CMEs) which is as expected for a time of low solar activity. If denser areas of solar wind are present in the field due to increased solar activity, the g-map would contain clusters of sources with heightened g-levels. It is clear to see, that the further out from the centre of the FOV the source is, it is more likely to show a g-level other than 1. This is due to the effect of noise as the sensitivity decreases further out in the primary beam, which is where the MWA is most sensitive too.

During the processing of the survey catalogue, a large number of sources at very low signal to noise (1\,$\sigma$) were kept. This was useful for the purposes of the survey, but due to the unreliability in the individual g-level measurements that such low signal to noise sources can create, leading to an inaccurate g-map, any source with a signal to noise ratio (S/N) of less than 2 was excluded from our analysis.

\section{Results} 
\label{sect:result}
Over the full search of 97 observations, more than 80\% of the g-maps resembled the one shown in Fig.\,\ref{fig:random}, with few extreme g-level sources or no obvious solar wind structures. The remaining $\sim$20\% have either an increased number of high g-level sources, both structured and random. We defer a more comprehensive analysis of the full data set to future work, and focus here on the most interesting structure detected. 

\subsection{Identification of a Moving Heliospheric Structure}
\label{sect:him}

\begin{figure*}[h]
    \centering
    \includegraphics[width=\textwidth]{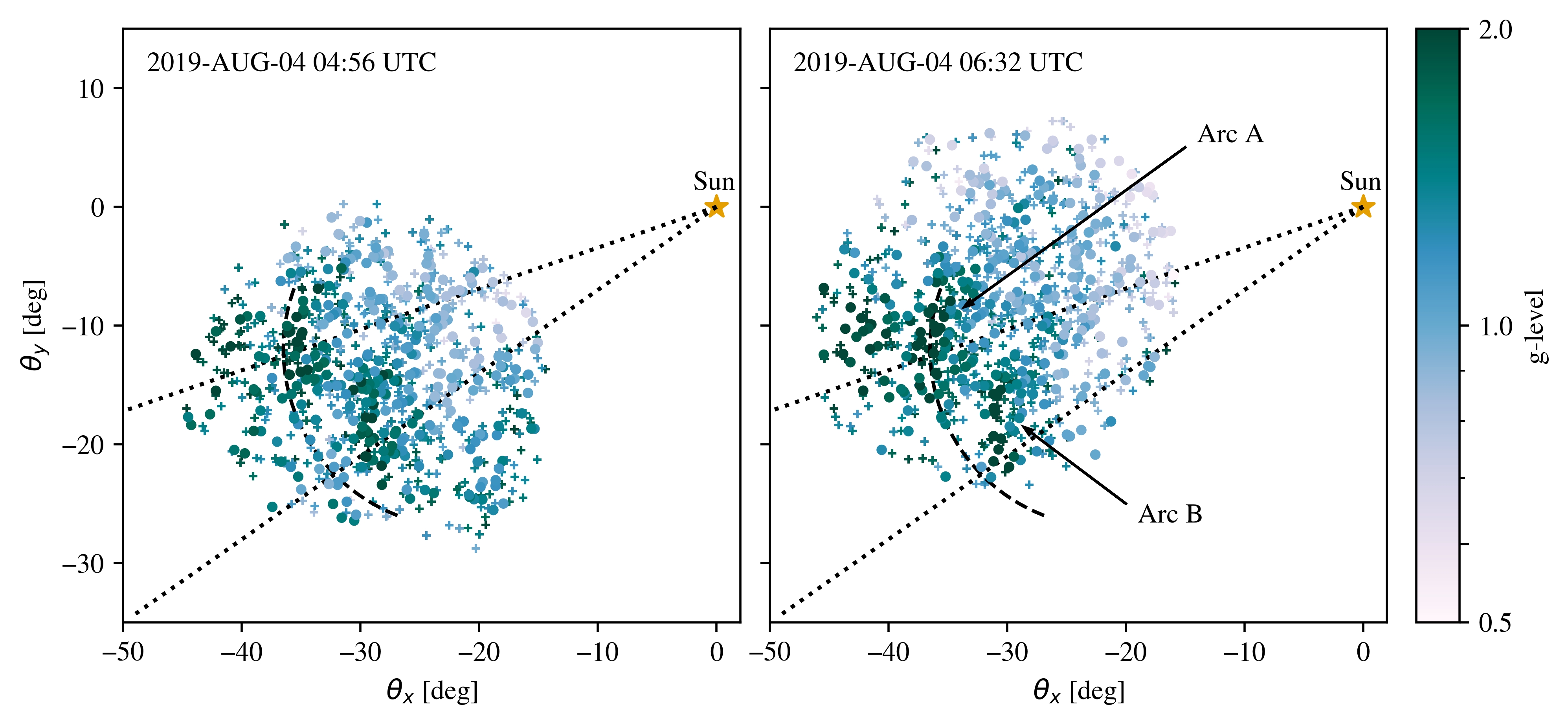}
    \caption{g-maps of MWA IPS observations in heliocentric coordinates separated by 1.5 hours in time, with g-level depicting the level of scintillation enhancement experienced by an individual source indicated with their colour. Left: Earlier observation taken at 2019-AUG-04 4:56 UTC with a pointing direction of SE. Right: Later observation taken at 2019-AUG-04 6:32 UTC with a pointing direction of E. 
    In both observations, three regions of enhanced scintillation can be identified; Arc A, Arc B, and a large shapeless mass on the very edge of the field.
    The dotted lines pass through the centre of each arc (the arc's position angle) as seen in the earlier observation. These dotted lines remain identical in the later observation, to act as a reference point in showing the radial movement of the feature away from the Sun. The dashed line in both figures is identical, and is identified as the leading edge of Arc A (as seen in the earlier observation). This shows further the radial movement of the structures away from the Sun.
    }
    \label{fig:gplots}
\end{figure*}

Our blind search revealed a handful of structures of which one event is particularly noteworthy, observed on 2019-AUG-04 at 04:56 UTC and 06:32 UTC. Three regions of enhanced scintillation are shown in Fig.\,\ref{fig:gplots}, and were found in two observations separated by $\sim$1.5 hours; two prominent arcs, one closer to the Sun (Arc B), and one further east away from the Sun (Arc A), alongside a large shapeless mass on the very edge of the field. Although the structures are irregular, which make measuring the exact morphology of the features difficult, each arc is about $5^\text{o}$ in width, or $\sim18$ solar radii wide, with the centre of the field being 0.5\,AU ($\sim107$ solar radii) from the Sun. 

This mass is not just on the edge of the field, it is also on the very edge of the primary beam, where the MWA is most sensitive too, while also being on the edge of the IPS survey coverage area. By being on the edge of the survey area, these particular sources within the mass were observed only a handful of times, compared to those further in being observed multiple times. This can cause uncertainties in the NSI of the sources, therefore increasing the uncertainty of their g-level. Further adding to these arguments, the mass has a high $\epsilon$, further increasing the uncertainty of the g-levels. For these reasons, this mass was excluded from the analysis.

Arc A and Arc B are slightly misaligned from each other, while still moving radially away from the Sun, with a $\phi$ of $\sim$$102^{\text{o}}$ and $\phi$ of $\sim$$115^{\text{o}}$ respectively. As evident from Fig.\,\ref{fig:gplots}, the structure appears to be moving through the FOV radially away from the Sun.

\subsection{Measuring the Angular Velocity}

\begin{figure*}[h!]
    \centering
    \includegraphics[width=\textwidth]{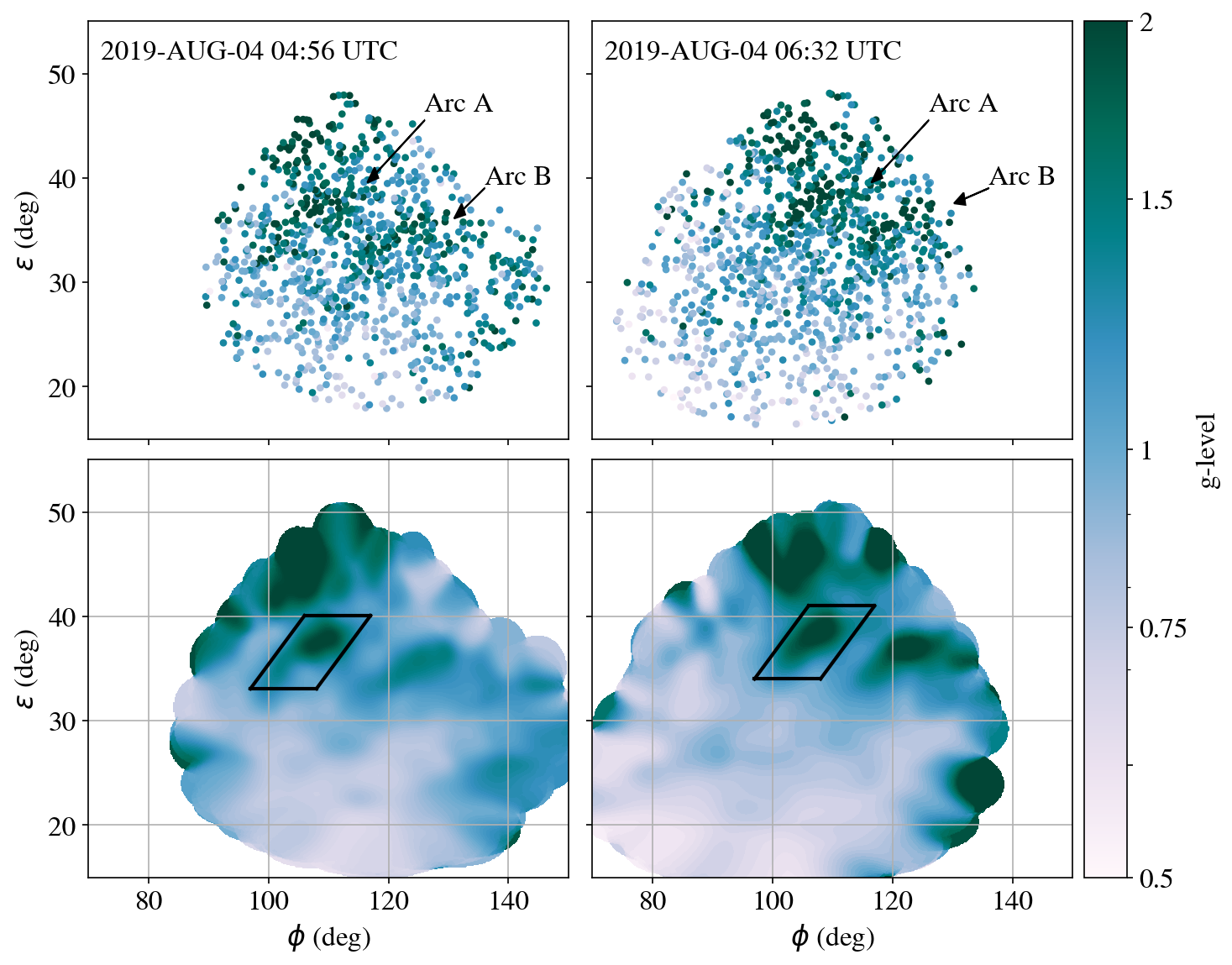}
    \caption{Top Row: Original g-maps for both observations in new radial heliocentric coordinate system, $\epsilon$ and $\phi$, where $\epsilon$ is degrees radially away from the Sun, and $\phi$ is the position angle. Both Arcs A and B are labelled. 
    Bottom Row: Smooth g-map for both observations with a pixel size of $0.2^{\text{o}}\times0.1^{\text{o}}$. 
    Arcs A and B are in the same location in the Bottom Row as in the Top Row. The black box on the reference observation (right) was drawn to define the enhanced scintillation area of Arc A. The black box on the target observation (left) is of the same size and shape, but has been shifted in down in $\epsilon$. 
    }
    \label{fig:blurs}
\end{figure*}

To make a measurement of the plane-of-sky velocity (i.e. the component of the true velocity perpendicular to the line of sight) of the detected heliospheric structure, we first interpolated the g-levels of all the sources onto a finer grid giving us a smooth g-map, as shown in Fig.\,\ref{fig:blurs}. As we expect the structure to move radially away from the Sun, we adopted a heliocentric coordinate system, as explained in Section\,\ref{sect:intro}. This reduces the problem to just one dimension, with change only in $\epsilon$ for both arcs. 

In Fig.\,\ref{fig:blurs}, we see both observations with their original g-map transferred to the new coordinate system on the top, with the bottom row showing this new interpolated map, with a pixel size of $0.2^{\text{o}}\times0.1^{\text{o}}$. The smoothed field was created by determining the g-level at any certain point using a radial basis function \cite<RBF,>{buhmann_2000} with a Gaussian form,
\begin{equation}
    \frac{\sum^{n}g_{n}w_{n}}{\sum^{n}w_{n}}, \text{ where } w_{n} = \text{exp}\left(-0.5\left(\frac{r}{r_{\text{o}}}\right)^2\right)\text{ for r $< 3^{\text{o}}$; 0 otherwise}
\end{equation}
Using this particular interpolation scheme with a radius of $3^{\text{o}}$ to search for nearby g-level measurements produced a smooth, completely defined g-map with no internal gaps in g-level, with both structures clearly visible in both observations

\begin{figure*}
    \centering
    \includegraphics[width=0.45\textwidth]{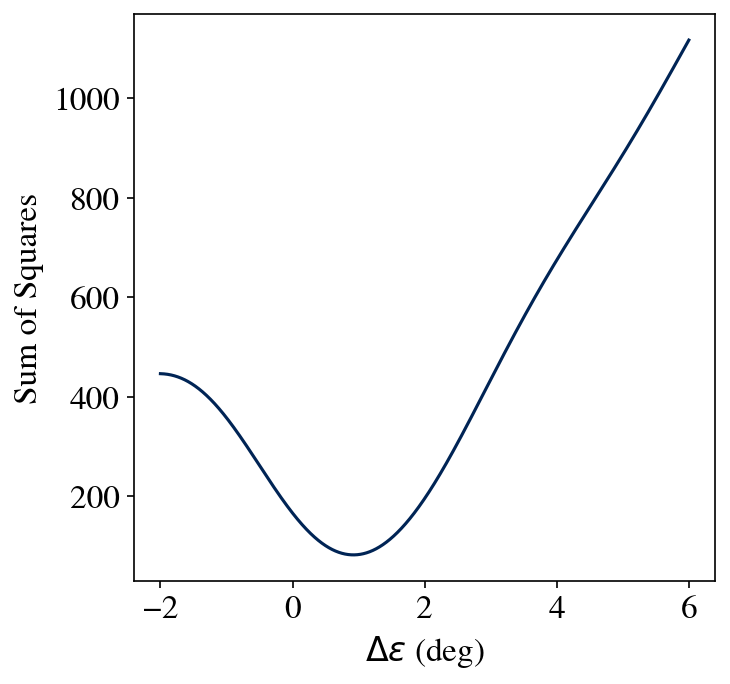}
    \caption{Sum of the differences in g-level squared between the boxes in the target and reference observations. This test was completed using the contributions of all sources with a defined g-level in the reference observation's field.}
    \label{fig:chi}
\end{figure*}

With fewer sources of lower S/N contributing to the g-level on the edges of the map, high g-level values can dominate and skew the map. This is of no concern to us as both Arcs A and B are well-defined within the centre of the g-map.
As the overall scintillation enhancement appears to be stronger in the later observation (on the right-hand side of Fig.\,\ref{fig:blurs}), this was used as the reference observation to define the location of Arc A by eye. A small box was defined to encompass the full arc, and the total g-level within the box was recorded. That same box was layered atop of the target observation (the earlier observation, on the left-hand side), and is allowed to move freely in $\epsilon$. A shift in $\epsilon$ of $-2^{\text{o}}$ corresponds to an angular velocity of $-1.38\,^{\text{o}}\text{hr}^{-1}$, meaning the structure would be moving backwards, and a shift of $6^{\text{o}}$ corresponds to a velocity of $4.11\,^{\text{o}}\text{hr}^{-1}$, equating to $3000\,\text{km}\,\text{s}^{-1}$ at 1\,AU, faster than any CME detected. By minimising the sum of the differences in g-level squared within the box between the two observations, shown in Fig.\,\ref{fig:chi}, the optimal shift in $\epsilon$ is determined. An estimated plane-of-sky velocity was found of $0.66\,^{\text{o}}\text{hr}^{-1}$. 

Although we have a strong prior that the only movement that this feature would exhibit would be radial this far from the Sun, there could be some movements present in $\phi$ (i.e. lateral changes). Running a similar analysis as was done for the radial velocity proved challenging due to the structures being extended in the radial direction, and once run, the jackknife tests indicated that, in contrast to the radial velocity, the result was highly dependent on which sources were included. 

It must also be noted, that this radial velocity estimation of Arc A depends on the assumption that this system only consists of a frozen screen moving in a radial direction through the field. This means that there are no changes within the screen itself between the two observations. This model holds true for Arc A, being well-defined and having a similar overall g-level in both observations, but this assumption does not hold for Arc B. Figure\,\ref{fig:blurs} clearly shows that in the earlier observation Arc B has a lower average g-level compared to the later observation. Although we are able to define the area of Arc B well in the later observation, as Arc B does not follow the assumed model, we are unable to obtain good constraints on Arc B’s movement. For this reason, we continued the plane-of-sky velocity estimation with only Arc A.

The same process was repeated in a series of jack-knife tests \cite{jackknife} to estimate the error on the plane-of-sky velocity. This involved each source in the field to be removed individually, the smooth g-map re-interpolated, and the new optimum $\epsilon$ shift to be found. Each of these shifts in $\epsilon$ were recorded, allowing two analyses to be done, first, a search for bias due to a single source dominating the smooth g-map, which none were found, and secondly, giving an estimate of the error of this velocity calculation. In total there were 1\,413 jack-knife tests, equal to the total number of sources in the reference field, above the designated signal to noise ratio. The jack-knife test gave an error of 22\%, with an adjusted final radial velocity of $0.66\pm0.147\,^{\text{o}}\text{hr}^{-1}$. Assuming a distance from Earth to the FOV of 1AU, it gives us a plane-of-sky velocity of $480\pm106\,\text{km}\text{s}^{-1}$. 
Extensive testing show that this result was insensitive to the size of the box or the exact form of the smoothing function.


\subsection{Exploring Possible Origins}


In order to determine the possible origin of the heliospheric structure, we undertook a systematic search of already catalogued solar events in the literature.
When a bright CME is detected by IPS observations, the g-level scintillation enhancements are expected to form arc-shaped structures, just as both \citeA{Morgan2023} and \citeA{TOKUMARU2013} found when studying large CMEs using the MWA and SWIFT, respectively. \citeA{TOKUMARU2013} explained it as compressed plasma associated with the leading edge of the interplanetary CME (ICME). These arc-shaped structures are similar to those seen in this observed scintillation enhancement, which encourages the idea that the structures we see originate from the Sun.
However, \citeA{Bisi2010} found that a dense compression region at the leading edge of a fast stream interacting with the trailing edge of a slow stream, such as in a stream interaction region, can also cause enhanced scintillation in IPS observations.
Given that these observations were taken close to solar minimum, when CMEs are relatively rare, but coronal holes closer to the equator are not \cite{Gopalswamy2022}, this alternative explanation is also worth considering.

Given a radial velocity, it is possible to find an estimated time of launch off of the Sun, assuming no acceleration is experienced by the structure. IPS is measured along a line-of-sight where it's highest sensitivity is at the closest approach to the Sun, the piercepoint. As we are observing $30^{\text{o}}$ away from the Sun, we can assume that the piercepoint, and therefore, the centre of the observation is at 107 solar radii. This can be assumed as the distance travelled by the observed heliospheric structure. Using both the plane-of-sky velocity and the distance travelled, it is estimated that the solar event that caused the structure would have been launched around 2019-AUG-02. In reality, solar wind events will accelerate or decelerate, therefore a reasonable launch window is from 2019-AUG-01 to 2019-AUG-03.
Along with the position angle for both arcs, mentioned in Section\,\ref{sect:him}, this information can be used to search through solar event catalogues as to find a possible progenitor on the solar surface. 

Various solar observatories, e.g. SOHO/LASCO \cite{brueckner1995} and STEREO/SECCHI \cite{howard2008}, catalogue major solar activity, specifically CMEs (and ICMEs). Alongside these catalogues there are additional catalogues of CMEs using alternative detection methods, some with computer generated catalogues, others with particular sorting techniques. A search of seven separate CME/ICME and solar event catalogues (listed in Section\,\ref{open}) was completed, with a known position angle, and an estimated velocity and launch time for this observed structure. No plausible match was found with all three criteria met. Even with less strict searches, such as excluding final velocity, there were still no plausible CME event filed in any of the catalogues searched.

\subsection{Coronagraph and EUVI Images}
The catalogues that were searched are all created using the white-light coronagraph images taken by either SOHO/LASCO or STEREO/SECCHI, therefore we also examined all white light coronagraph difference images made by SOHO/LASCO C2 and C3 detectors over the period of estimated launch, to conclude that there was no obvious activity caused by a CME or solar flare event.
However, during the estimated time of launch of the observed structure, STEREO-A was facing the eastern limb of the Sun (see Figure\,\ref{fig:target} for the relative positions). In the EUVI images made by STEREO-A during the estimated launch time, there are two coronal holes that are visible. The first is an equatorial coronal hole, which created a high-speed stream that later impacted STEREO-A \cite{hss}, while the second is a low-latitude coronal hole, matching the position angle of our observed feature. The high-speed stream from the equatorial coronal hole reached a peak solar wind speed of 460\,$\text{kms}^{-1}$. Although from a different coronal hole, this supports the hypothesis that the structure that we observed could be a stream interaction region (SIR). As noted above such events have been detected with IPS techniques in the past \cite{Breen1998, Bisi2010}.



\section{Discussion}
\label{sect:discuss}

With a well calculated estimation of the plane-of-sky velocity and a positional analysis completed, it is clear that the observed structures have solar origins. Although some preliminary work in deciphering the exact solar origins of this observed structure has been done, by looking at both white-light coronagraph images as well as STEREO EUVI images, the exact nature of the structures remains unclear. Whether this structure originates from a stream interaction region, or possibly a small, undetected CME impacting a SIR, or another solar event, can not be differentiated at this time. 

\subsection{Implications for Heliospheric Monitoring}

IPS observations taken by the MWA can provide a unique viewpoint of the heliosphere that many other solar probes and IPS stations are not able to provide. Using our MWA IPS observations, we were able to detect a heliospheric structure that would have otherwise gone undetected. \citeA{Braga2022} state that the region between the solar corona and 1AU is not probed to the fullest extent. Coronagraph imaging has a limited field of view, with the majority of current instruments aimed relatively close to the Sun, where the largest fields of view reach only about 8 degrees ($\sim$32 solar radii) away from the Sun. As shown, the MWA is able to probe much further out into interplanetary space, monitoring how space weather events might evolve as they move with an unprecedented density of detected scintillating sources.

With the MWA's ability to sample any region of space surrounding the Sun, it possesses the ability to remotely sense solar wind on the Eastern limb of the Sun. Activity in this region originated on parts of the solar surface that have not be viewable from the Sun-Earth line for upwards of 13 days, so data for this part of the heliosphere can be particularly scarce. This lack of information can lead to uncertainties in models, where magnetic field coverage is limited in certain areas of the heliosphere \cite{Jin2022}.

Our ability to infer the angular velocity with MWA IPS on the sky provides independent information on the velocity of detected structures. It is generalised enough that an angular velocity can be found for any structure in IPS data, whether it be a stream interaction region or CME. 
Work carried out by the majority of the IPS space weather community is measuring and tracking the radial speed of CMEs and ICMEs, and as discussed by \citeA{Iju2013} and \citeA{Iwai2021}, it is the use of multiple IPS stations and a variety of techniques that can give the best interpretation of the solar wind. Our work is very complementary to multi-station IPS undertaken by ISEE \cite{TOKUMARU2013} and single-station power spectrum fitting \cite{Chang2019} with LOFAR.

\subsection{Future Work}
\label{sect:future}
As previously stated in Section\,\ref{sect:discuss}, the nature of the observed structures as well as their exact solar origins are unclear. The use of MHD or full-scale solar wind simulations may be useful in aiding the interpretation of the origins and evolution of this event. However, such modelling is beyond the scope of this initial discovery publication, and we defer this more in-depth simulation analysis for a future publication.

Since the completion of this work, we now have IPS data covering 20 months between April 2020 and March 2023, with the observing periods being 2020-APR-13 to 2021-JAN-28, 2022-JUN-15 to 2022-OCT-20, and 2022-OCT-29 to 2023-MAR-20. These observations were taken in survey mode as depicted in Section\,\ref{sect:observations} (similar target fields to Fig.\,\ref{fig:target}), and we intend to analyse a significant subset of these data in order to continue the search for interesting heliospheric activity.
This initial work relied on visual inspection of all the processed g-maps as a search for any interesting observations or features. Since it was still unknown whether anything of interest would be present in the data, this visual inspection method sufficed. For future work with MWA IPS data, we plan on implementing a systematic process that would flag possible candidate observations for further study.

The Australian SKA Pathfinder \cite<ASKAP,>{askap} is a higher frequency radio telescope dish array which is co-located with the MWA. As shown by \citeA{Chhetri2022}, IPS measurements can be made using a similar scheme (assuming the same number of pointings and observing time as in previous sections) probing $5-20^{\text{o}}$ from the Sun. This would allow us to make almost continuous observations from $\sim$20 to over 100 solar radii. Where ASKAP reaches its limits, the MWA takes over. 

In the future there is is possibility of doing triggered MWA observations of a known CME. As the MWA is probing far into the heliosphere, even a very fast CME will take a number of hours to reach the MWA's FOV, this gives us enough time to schedule observations. As long as the Sun is above the horizon, we are able to take survey mode observations over the full day ($\sim$8 hour period will result in $\sim$48 observations), probing in particular the location of the CME. The CME would take several hours to leave the MWA's FOV, meaning there is a high likelihood of having more than two observations of the structure. With more observations we can measure the velocity with more precision, and/or estimate any acceleration.

\subsection{Conclusions}
\label{sect:conclude}
With two interplanetary scintillation observations separated by 1.5 hours, taken by the Murchison Widefield Array during mid-2019 close to sunspot minimum, we observed a moving heliospheric structure using the high density of IPS sources in the field-of-view ($\sim$700 detected sources in $900\,\text{deg}^{2}$). We observe g-levels greater than 1.5, implying highly enhanced levels of scintillation caused by increased density within the solar wind. As two individual observations were made of the same structure, a radial plane-of-sky velocity was able to be inferred of $480\pm106$\,$\text{kms}^{-1}$. This provides an excellent demonstration of the benefits of the MWA's large field of view which allows for simultaneous observations of a large number of compact sources and their IPS characteristics.

After comparisons with seven separate CME and ICME catalogues, as well as white light coronagraph difference images, we conclude that this heliospheric structure was not associated with a Coronal Mass Ejection. With the link of stream interaction regions (SIRs) having a solar cycle dependency during the declining phase, and images from STEREO-A of a coronal hole on the Sun at low-latitudes corresponding to the position angle of our observed heliospheric structure, we hypothesise that our structure is an IPS observation of a SIR, in what is considered far interplanetary space.

The MWA is able to probe the interplanetary space where current measurements are sparse, especially regions far from the corona out to 1AU. With a large density of IPS sources per observation, compared to other current IPS stations, the MWA has a unique capability of providing important information, such as the structural evolution, of the solar wind over a large region, that is unable to be obtained at high cadence from other techniques and instruments. 

\section{Coordinate Definitions}
\label{sect:def}
\begin{acronyms}
  \acro{Helioprojective Radial Coordinates}
  To describe our observations used in this work we use a helioprojective coordinate system centred on the observer (the Earth). Any observing direction can be parameterised by $\epsilon$ and $\phi$, where $\epsilon$ is the elongation from the Sun, while $\phi$ is the position angle measured from the Sun's North pole (projected into the plane of sky of the observer) through East.
  \acro{HCC - Heliocentric Cartesian Coordinates}
  A coordinate system in the heliocentric system which is observer-based. The origin is the center of the Sun. The Z-axis is aligned with the Sun-observer line. The Y-axis is aligned with the component of the vector to the Sun's north pole that is perpendicular to the Z-axis.
  \acro{HPC - Helioprojective Cartesian Coordinates}
  A coordinate frame which is observer-based. The origin is location of the observer (the Earth). $\theta_x$ is the angle relative to the plane containing the Sun-Earth line and the Sun's rotation axis. $\theta_y$ is the angle relative to the Sun's equatorial plane. This coordinate system and the earlier helioprojective radial coordinate system are related as so; $\theta_x=\epsilon\sin\phi$, and $\theta_y=\epsilon\cos\phi$. 
\end{acronyms}




\section{Open Research}
\label{open}
MWA data is available from the MWA All-Sky Virtual Observatory \cite{ASVO}, and for this work was accessed via giant-squid \cite{giant}, which is an alternative MWA ASVO client. For access to the data stored in this archive, registration is required. At the time of writing, the observations used in this paper are public, and can be identified by their GPS start times (example g-map: 1247722256, detected structure: 1248929800 and 1248935560) which serve as unique identifiers of these observations within the MWA archive. All the IPS observations described in \citeA{ipssurvey} are also archived, under project code \textsc{D0011}.

Coronal Mass Ejection public catalogues that were queried for this work are as follows; SOHO LASCO CME Catalog \cite{Gopalswamy2009}, STEREO COR1 CME Catalog \cite{stereo-cor1}, CACTus LASCO C2/C3 catalogue and COR2 catalogue \cite{Robbrecht2004, Robbrecht2009}, SEEDS LASCO C2 catalogue \cite{seeds}, SEEDS SECCHI COR2 catalogue \cite{secchi}, and WIND ICME Catalogue \cite{wind}.

This research used version 4.0.2 \cite{sunpy} of the SunPy open source software package \cite{sunpy_community2020} for coordinate conversions.


\acknowledgments
This scientific work makes use of Inyarrimanha Ilgari Bundara, the Murchison Radio-astronomy Observatory operated by CSIRO.
We acknowledge the Wajarri Yamatji people as the traditional owners of the Observatory site.
Support for the operation of the MWA is provided by the Australian Government (NCRIS), under a contract to Curtin University administered by Astronomy Australia Limited.
We acknowledge the Pawsey Supercomputing Centre which is supported by the Western Australian and Australian Governments.
A.W was supported by an Australian Government Research Training Program (RTP) Stipend and RTP Fee-Offset Scholarship.

\bibliography{references}

%
%
%
%
%

\end{document}